\documentclass[journal=jpccck,layout=twocolumn]{achemso}  

\usepackage{chemformula} 
\usepackage[T1]{fontenc} 
\usepackage{xcolor}
\definecolor{darkgreen}{HTML}{008000}
\usepackage{siunitx}

\makeatletter
\def\acs@author@fnsymbol#1{}   
\makeatother

\author{Vincent Ballenegger}
\email{vincent.ballenegger@univ-fcomte.fr}
\affiliation[affiliation1]
{Institut UTINAM, Univ. Bourgogne-Franche-Comté, UMR CNRS 6213,\\ 16, route de Gray, 25030 Besançon Cedex, France}

\title[running title]
  {Cage occupancies in nitrogen clathrate hydrates\\ from Monte Carlo simulations }

\abbreviations{IR,NMR,UV}

\begin{document}

\begin{tocentry}

Some journals require a graphical entry for the Table of Contents.
This should be laid out ``print ready'' so that the sizing of the
text is correct.

Inside the \texttt{tocentry} environment, the font used is Helvetica
8\,pt, as required by \emph{Journal of the American Chemical
Society}.

The surrounding frame is 9\,cm by 3.5\,cm, which is the maximum
permitted for  \emph{Journal of the American Chemical Society}
graphical table of content entries. The box will not resize if the
content is too big: instead it will overflow the edge of the box.

This box and the associated title will always be printed on a
separate page at the end of the document.

\end{tocentry}

\begin{abstract}
Comparisons of Gibbs ensemble Monte Carlo simulations with experimental data for the cage occupancies in \ch{N2} clathrate hydrates are performed to assess the accuracy of such simulations, to refine the effective potentials employed, and to help interpret recently measured large cage over small cage occupancy ratios [Petuya et al., J. Phys. Chem. C 122, 566 (2018)]. Different sets of interaction potentials for \ch{N2-N2}, \ch{N2-H2O} and \ch{H2O-H2O} interactions are considered. Some of them fail to reproduce the known experimental fact that some large cages are doubly occupied at 273~K and high pressures. The best agreement between simulations and experiments is obtained when using a new \ch{N-O} interaction potential derived in this work by averaging an ab-initio potential energy surface for the \ch{N2-H2O} dimer.
\end{abstract}

\section{Introduction}

A gas clathrate hydrate is a form of crystalline ice where the water molecules form a solid network with small and large cavities that encapsulate guests~\cite{Sloan2007}. This molecular structure is stable at high pressure and/or low temperature. Methane (clathrate) hydrates have been found in natural gas pipelines, deep sea sediment and permafrost regions. They are considered to be an enormous possible future energy resource~\cite{Sloan2007}. Their exploitation could be combined with the sequestration of carbon dioxide, in the form of \ch{CO2} hydrate on the seafloor, to help keeping the concentration of this global warming gas under control\cite{Ohgaki1994}. Nitrogen molecules were found to act as a promoter in the \ch{N2}–assisted \ch{CH4-CO2} exchange reaction\cite{Park2006}.  Since flue gas mixtures contain a mixture of \ch{N2} and \ch{CO2}, a deep understanding of mixed \ch{N2-CO2} clathrates, and also pure \ch{N2} and \ch{CO2} clathrates as reference systems, is important for this technology. Air clathrates, containing \ch{N2}, \ch{O2} and other trace gases, are found on Earth below ice caps at a depth of \SI{1000}{m} or more. They are of particular interest as they contains relics of our ancient atmosphere up to 420 kyr back. \ch{N2} clathrates may also be present in comets and in some planets or their satellites of our solar system\cite{Thomas2008,Mousis2010}.

There exist 3 clathrate structures at low pressures: two cubic ones (sI, sII) and a hexagonal one (sH)\cite{Sloan2007}. \ch{N2} clathrates crystallize in a cubic structure: the structure I is kinetically favoured, while the sII is the thermodynamically stable structure if the pressure is not too high (a transition to sH is observed at the high pressure 0.85 GPa)\cite{Petuya2018,Loveday2003,Sasaki2003}. The unit cell of the sI crystal contains 46 water molecules forming 8 cages: 2 small and 6 large ones.  The unit cell of sII contains 136 water molecules forming 24 cages: 16 small cages and 8 large cages (these large cages are slightly bigger than in sI). Structure II is promoted by the ability of the small molecule \ch{N2} to fill in the numerous small cages of this structure.

The cage occupancies $\theta_{\rm S}$ and $\theta_{\rm L}$ of small (S), respectively large (L), cages in a clathrate is an important quantity for practical applications: they determine the gas storage capacity (see e.g. Ref.~\cite{Brumby2019} for hydrogen storage) and intervene also in the calculation of phase equilibria involving clathrates\cite{Sloan2007,Lasich2014}. Occupancies in a nitrogen clathrate have been measured, at \SI{273}{K} and pressures up to \SI{1000}{bar} using powder neutron diffraction and Rietveld refinement by Chazallon and Kuhs~\cite{Chazallon2002} for both sI and sII. They discovered that some large cages in sII can be doubly occupied at pressures above $\approx$~\SI{300}{bar}. An additional measurement at \SI{1093}{bar} and \SI{268.2}{K} in a sI nitrogen clathrate was performed by Qin and Kuhs\cite{Qin2015}; it also shows some double occupancy of the large cages: $\theta_{\rm L} = 111.9 \pm 0.8$\% while $\theta_{\rm S} = 98\%$. Yet another measurement in a sII clathrate at \SI{150}{bar} and at a lower temperature \SI{258.15}{K} gave: $\theta_{\rm L} = 121.8 \pm 0.6$\% and $\theta_{\rm S} = 96.7 \pm 0.3\%$\cite{Hansen2016}. Occupancy ratios $\theta_{\rm L} / \theta_{\rm S}$ have moreover been measured at pressures up to 200 bar and for temperatures down to \SI{150}{K} by Petuya et al\cite{Petuya2018}. The latter authors used another technique, Raman spectroscopy, which does not provide separately the occupancies of large and small cages, but only their ratio.

The purpose of the present paper is to perform Monte Carlo simulations to determine the occupancy of cages in \ch{N2} clathrates at thermodynamical equilibrium for the same conditions as in the previous experiments. This will enable us to assess the agreement between simulations and experiments for a wide range of conditions covering both single and multiple occupancy regimes for the cages, and to help interpret the recently measured occupancy ratios\cite{Petuya2018} in terms of separate occupancies $\theta_{\rm L}$ and $\theta_{\rm S}$. Notice that similar comparisons but for the simpler methane clathrate, which forms in structure~I and which does not show any multiple occupancy of cages, have been done by several authors\cite{Henley2015,Papadimitriou2016,Brumby2016}. 

Only few simulations results for a \ch{N2} clathrate can be found in the literature\cite{Horikawa1997,Klapproth1999,vanKlaveren2001,vanKlaveren2001b,vanKlaveren2002,Patt2018}. Horikawa et al., and later van Klaveren et al., have studied the dynamical behavior of encaged \ch{N2} molecules using molecular dynamics (MD)\cite{Horikawa1997,vanKlaveren2002}. van Klaveren et al. have shown also that a \ch{N2} hydrate with sII at high pressure remains {\it mechanically} stable even for the case of a full double occupancy of the large cages.\cite{vanKlaveren2001}. In those MD simulations, the number of guest \ch{N2} molecules is fixed beforehand and no attempt was made to quantify the {\it thermodynamical} stability of a phase with full or partial double occupation. The Monte Carlo (MC) method overcomes this difficulty because MC moves with insertion/deletion of guest molecules allow the average cage occupancy of the simulated clathrate to converge automatically towards its value at thermodynamical equilibrium. A grand-canonical Monte Carlo (GCMC) calculation of cage filling in a \ch{N2} clathrate has been performed by Klapproth et al.\cite{Klapproth1999}, who found an unsatisfactory agreement: simulations strongly overestimated the occupancy when using the simple point charge (SPC) water model.\footnote{This overestimation was probably due in part to their assumption of a rigid water framework.
}
Patt et al. performed GCMC simulations of \ch{N2}, \ch{CO} and mixed \ch{N2-CO} clathrates but only at low temperatures ($50$, $100$ and \SI{150}{K}) and they did not compare quantitatively with experimental data\cite{Patt2018}. Their calculations suggest that experiments performed on the single-guest \ch{N2} and \ch{CO} clathrates might be sufficient to get information on the corresponding mixed clathrates.

In the present work, the Gibbs ensemble Monte Carlo (GEMC) technique is employed because of its convenience: the experimental pressure and temperature are direct input parameters for the simulation and the volume of the simulated clathrate adjusts itself automatically to the imposed pressure. This can be contrasted with grand-canonical MC simulations where the volume is fixed (and must hence be adjusted manually when varying the pressure) and where the pressure of the gas is not controlled directly but must be deduced from its fugacity by using an appropriate equation of state.

\section{Computational details}

Isobaric-isothermal Gibbs ensemble Monte Carlo simulations were performed with the open-source MCCCS Towhee simulation program\cite{Towhee} (version 7.2.0) to compute the cage occupancies $\theta_{\rm S}$ and $\theta_{\rm L}$ for pressures and temperatures where experimental data points are available. In this ensemble two simulation boxes are held at the same constant pressure $P$ and temperature $T$, while the total number of water and nitrogen molecules across both boxes, is also constant. One box represents the \ch{N2} clathrate crystal and the other the nitrogen gas phase. Some additional GCMC simulations were performed with the open-source DL\_MONTE-2 simulation code\cite{DLMONTE}; their results were consistent with those of Towhee.

\subsection{Molecular models}	\label{Sct2.1}
A classical description of the interaction potentials between water and guest molecules has been used. Short-range and dispersion interactions were computed as a sum of pairwise Lennard-Jones (LJ) or Buckingham (also known as Exp-6) interactions between interaction sites located on the \ch{H2O} and \ch{N2} molecules. Coulomb interactions between point partial charges were computed using the Ewald summation technique. 
Different force fields have been considered to gauge the sensitivity of the predicted cage occupancies on the choice of the force field.
The water molecules have been represented by the TIP4P-Ew\cite{TIP4PEw} and the TIP4P/Ice models\cite{TIP4Pice}.

To properly describe double occupancy of cages in a \ch{N2} clathrate, a highly accurate model for the repulsive interaction between two \ch{N2} molecules is required because they are in close contact when occupying the same cage. 
Two models were used to represent the \ch{N2} molecule: the model of Potoff et al.\cite{Potoff2001} which is included in the TraPPE force field, and the model of Etters et al.\cite{Etters1986}. The Potoff model is a 3-site model with a partial charge $q=-0.482$~a.u. on each N atom (which are separated by a distance $1.1$~\AA) and a partial charge $-2q$  on the center of mass, and with a LJ interaction associated with each \ch{N} atom. Its parameters (see Table~\ref{Table0}) are very close to the model X1 of Murthy et al.\cite{Murthy1983} that was used in the simulation study by Klapproth\cite{Klapproth1999}: the geometric and LJ parameters are almost identical, but the latter model differs by the fact that it features no point charge but a point quadrupole of strength $\SI{-3.91e-40}{C.m^2}$ located on the center of mass (that strength is considered to be an adjustable parameter in the X1 model and results from a best fit to a wide range of properties). The Etters model involves 6 interaction sites: a repulsion-dispersion interaction site on each N atom and 4 Coulomb interaction sites along the axis of the molecule at the positions $\pm 0.847$ and $\pm 1.044$\,\AA, i.e. further away than the N atoms which are at positions $\pm \SI{0.547}{}$\,\AA. 
\begin{table}[b]
\caption{Parameters for \ch{N-N} interactions}
\label{Table0}
\begin{tabular}{ccccc}
{\bf Model} & {\bf Type} &   $\sigma$ (\AA) & $\epsilon/k_{\rm B}$ (K) & $\gamma$\\
\hline
TraPPE\cite{Potoff2001} & LJ & 3.31 & 36.0 & -\\ 
Etters et al.\cite{Etters1986}  & Exp-6 & 3.3779 & 32.8519 & 13.1946 $^{(a)}$ \\
& Exp-6  & 3.3269 & 39.7347 & 15.0755 $^{(b)}$ \\
\hline
\multicolumn{5}{l}{
{\footnotesize
$^{(a)}$ if $r < 3.0102$~\AA;\quad  $^{(b)}$ if $r > 3.4495$~\AA;\quad  see Ref.\cite{Etters1986} otherwise }
}
\end{tabular}
\end{table}
Similarly to van Klaveren et al., we have modified slightly the electrostatic description to reduce the number of interaction sites to 3 and to match the experimentally observed quadrupole moment [\SI{-4.7e-40}{C.m^2}]. 
The repulsion-dispersion interaction potential in the Etters model was obtained from a refitting of ab initio calculations to better match experimental data for the second virial coefficient and for condensed phases of \ch{N2}. It is given by a Buckingham interaction at short distances, another Buckingham interaction at large distances, and a quartic spline fit at intermediate distances (see Table~\ref{Table0} and Ref.~\cite{Etters1986} for the parameters). This potential is less repulsive than the original ab-initio potential for an isolated pair of molecules: this softening was found to be necessary to accurately describe condensed phases.\cite{Etters1986}
\begin{figure}
\includegraphics[scale=0.57]{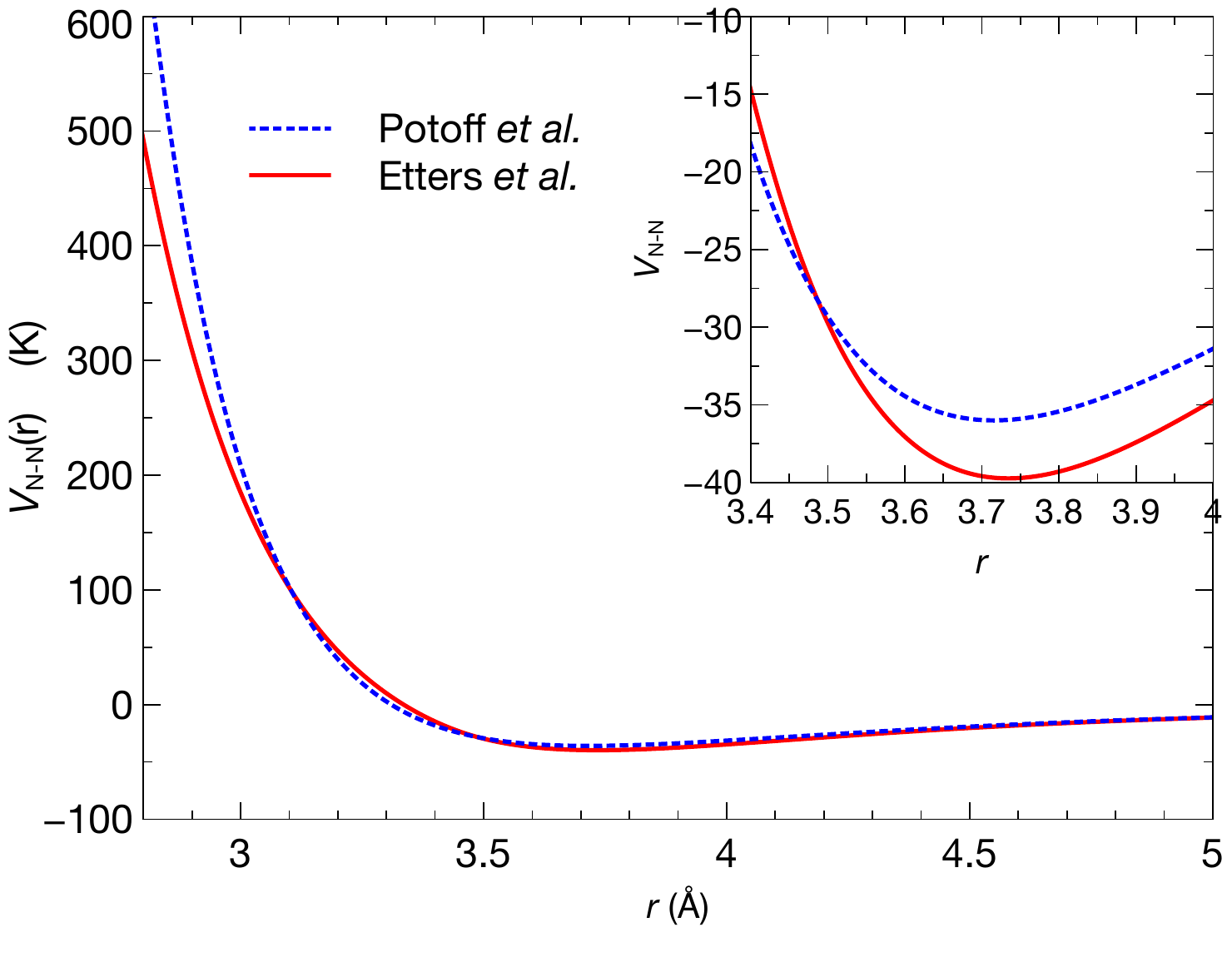}
\caption{\small The N site - N site interaction potential according to two different force fields: Etters et al.\cite{Etters1986} (red line) and TraPPE\cite{Potoff2001}. A zoom on the region around the potential minimum is shown in the inset.}
\label{Fig1NN}
\end{figure}
The nitrogen interaction potentials of Potoff {\it et al.} and of Etters {\it et al.} are compared in Fig.~\ref{Fig1NN}. The Etters potential is less repulsive at short distances and more attractive near the well's minimum.

The \ch{N2-H2O} interaction is particularly important in this work because it determines the available cage volume and the potential energy of the guest molecules inside the cages. van Klaveren {\it et al.} have shown that the preferred parallel orientation of two \ch{N2} molecules in a large cage is dictated by this interaction. In previous simulations on doubly-occupied \ch{N2} clathrates\cite{Klapproth1999,vanKlaveren2001,vanKlaveren2001b,vanKlaveren2002}, the \ch{N-O} interaction was described by a LJ potential, with parameters obtained via the Lorentz-Berthelot mixing rules. The LJ parameters used in Ref.\cite{vanKlaveren2001} originate for instance from Ref.\cite{Somasundaram1999} where they were calculated by combining parameters of the SPC/E water model with those of a \ch{N-N} interaction. The importance of deviations from LB mixing rules in the modelling of clathrates has been examined recently for various gases in the framework of the van der Waals-Platteuw model\cite{Papadimitriou2014b}, and also with Monte Carlo simulations in the case of argon hydrates\cite{Papadimitriou2014a}.

To refine the \ch{N-O} interaction, we have considered the full potential energy surface of the dimer \ch{N2-H2O} determined by Tulegenov et al.\cite{Tulegenov2007}. They have calculated \si{12228} energies for different intermolecular configurations with a post-Hartree-Fock method and have proposed a fit of the energy surface to a site-site interaction of the form
\begin{equation}	\label{Tul}
E_{\rm N2-H2O} \approx  \sum_{i={\color{black}\rm N_1,N_2}} \sum_{j={\color{black}\rm O, H_1, H_2}} \sum_{n=2}^{12} \sum_{\Lambda=0}^3 \, \displaystyle {\rm C}_{{{\color{black}i},{\color{black}j},n,\Lambda}} \, \frac{S_{i,j,\Lambda}(\Omega_{ij})}{R_{ij}^n}
\end{equation}
where $R_{ij}$ are intermolecular site-site distances and where the orientation function $S$ introduces anisotropy into the atom-atom interactions. This fit holds when the intermolecular distance $R$ is smaller than 12 Bohr radius (a different fit was proposed for distances larger than $14\,a_{\rm B}$). Eq.~\eqref{Tul} contains all contributions to the interaction energy: exchange-repulsion, dispersion and electrostatic energy, including induction energy. 
Since the proposed fit contains too many coefficients (564 in total) to be used in this form in our Monte Carlo simulations, we have deduced an average interaction potential $V_{\ch{N-O}}(r)$ by averaging over dimer configurations for a fixed distance $r$ between a \ch{N} and \ch{O} atom. In this calculation, the interaction \eqref{Tul} is first split into separate interactions with each of the two sites N$_1$ and N$_2$ of the nitrogen molecule. The interactions with the H$_1$ and H$_2$ sites are added to the \ch{N-O} interaction, i.e. the sum over $j$ in eq.~\eqref{Tul} is kept as is, to avoid picking up large electrostatic contributions. If the molecules were fully rigid and non-polarizable, all electrostatic contributions would cancel out in this averaging procedure when the two molecules are not touching each other (because the electrostatic potential outside a charged spherical shell is independent of its radius). The potential $V_{\ch{N-O}}(r)$ arises mainly from the exchange-repulsion and dispersion energies, but it does include also contributions from the induction energy.
 We have checked that the same average interaction $V_{\ch{N-O}}(r)$ is obtained, near the potential well minimum and also further away, when subtracting to $E_{\rm N2-H2O}$  the electrostatic interactions computed with the partial charges defined in the water and nitrogen molecular models. 
 
\begin{figure}
\includegraphics[scale=0.57]{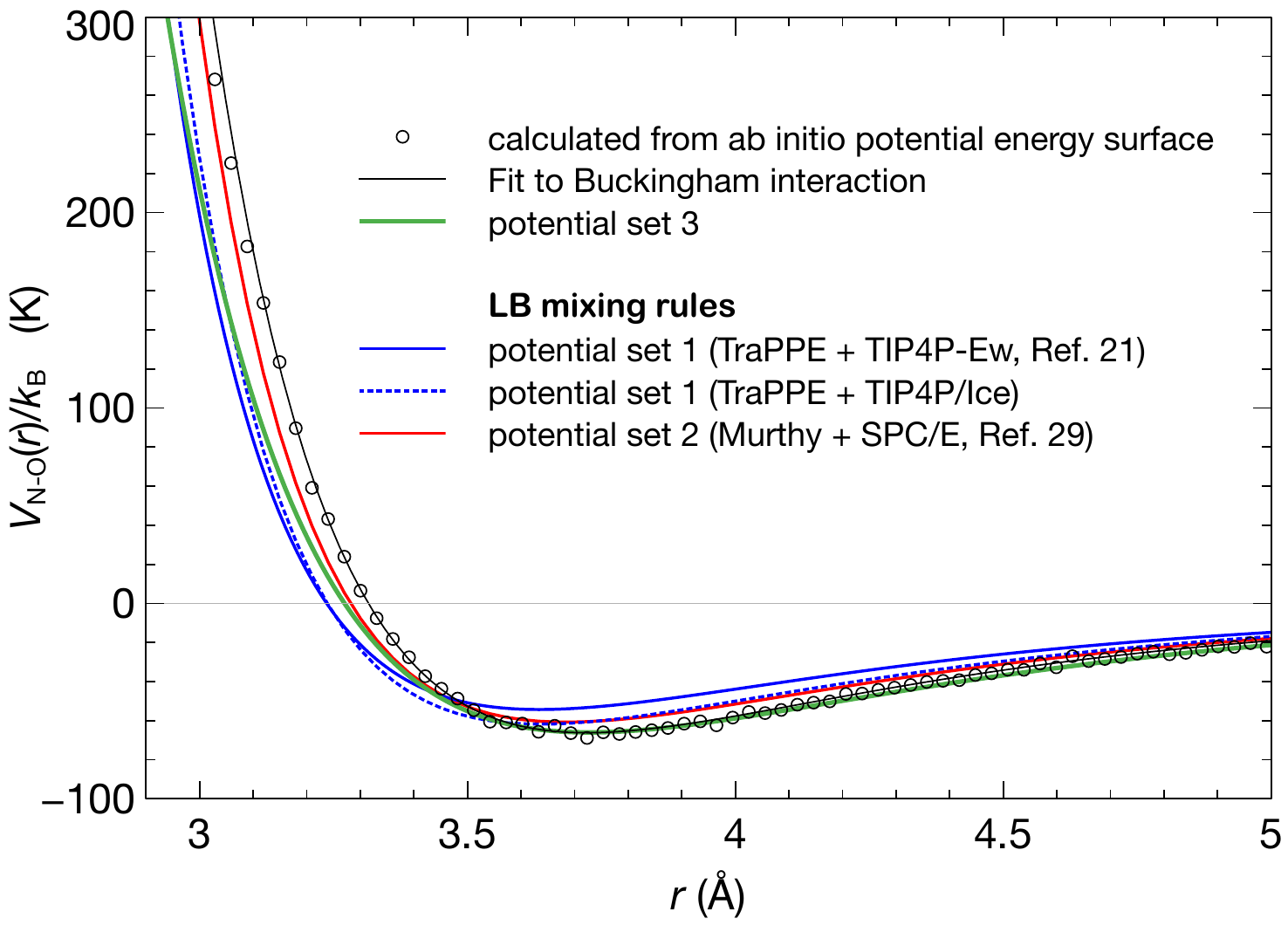}
\caption{\small The N site - O site interaction potential calculated from the ab-initio potential energy surface of Tulegenov et al.~\cite{Tulegenov2007} (circles) and its fit to an Exp-6 interaction (black line). The Exp-6 and the Lennard-Jones potentials listed in Table~\ref{Table1} are also shown (see color key in plot).
}
\label{Fig2}
\end{figure}

The potential $V_{\ch{N-O}}(r)$ that results from the previous averaging procedure is shown in Fig.~\ref{Fig2}. Fitting this potential to a Buckingham interaction
\begin{equation}	\label{Buck}
V_{\rm N-O}^{\rm Buck}(r) = \epsilon \left[ \frac{6}{\gamma-6} e^{-\gamma \big(\frac{r}{r_m} - 1\big)} - \frac{\gamma}{\gamma-6} \Big(\frac{r}{r_m}\Big)^{-6} \right],
\end{equation}
where $r_m = 2^{1/6}\sigma$, provides the parameters $\epsilon=66.2$~K, $\sigma=3.32$~\AA\ and $\gamma=14.3$.
The potentials obtained by using the Lorentz-Berthelot mixing rules (blue and red curves in Fig.~\ref{Fig2}) are quite close to this potential $V_{\rm N-O}^{\rm Buck}(r)$. It is somewhat more repulsive at short distance, and more attractive at larger distances. The depth of the potential well is about 9\% more attractive than the one in Refs.~\cite{Somasundaram1999,vanKlaveren2001} (red curve). The depth of the potential $V_{\rm N-O}^{\rm Buck}(r)$, deduced from the ab-initio potential energy surface for a dimer, should be accurate, but this potential is likely too repulsive at short distances when applied to a condensed phase, similarly to the case of the \ch{N-N} interaction.  We introduce therefore a softer potential (green curve in Fig.~\ref{Fig2}) by setting $\gamma$ to 12.2, a value obtained by fitting the \ch{N-O} potential of Ref.~\cite{Patt2018} (blue curve) to a Buckingham interaction. This provides the parameters listed in the first column of Table~\ref{Table1}. The remaining columns show parameters used in previous studies: Ref.~\cite{Patt2018} for the case of potential 1 with TIP4P-Ew water and Refs.\cite{Somasundaram1999,vanKlaveren2001,vanKlaveren2001b,vanKlaveren2002} for the case of potential~2.
\begin{table}[b]
\caption{Parameters for \ch{N-O} interactions}
\label{Table1}
\mbox{}\hskip -1.5cm \mbox{}
\begin{tabular}{c|ccc}
{\bf Potential} & {\bf \color{darkgreen}3} & {\bf \color{red}2} & {\bf \color{blue}1}\\
& {\small  This work} & {\small Refs. \cite{Somasundaram1999,vanKlaveren2001,vanKlaveren2001b,vanKlaveren2002}} & {\small TraPPE with} \\
& & & {\small TIP4P-Ew or -Ice}\\
\hline
Type & Exp-6 & LJ & LJ \\  
$\epsilon$ (K) & 66.2 & 60.74  & 54.30 / 61.80\\
$\sigma$ (\AA) & 3.32 & 3.282  & 3.237 / 3.238\\
$\gamma$ & 12.2 & - & - \\
\hline
\end{tabular}
\end{table}

The potentials sets 1, 2 and 3 involve the potentials 1, 2 and 3 for the \ch{N-O} interactions, respectively. In the potential set 1, the \ch{N2-N2} interactions are given by the TraPPE force field, while they are given by the Etters model in the potential sets 2 and 3. No standard mixing rule exist for combining the parameters of the Etters model for \ch{N2} with those of a water model. In the potential sets 2 and 3, the \ch{N-O} repulsion-dispersion interactions are those listed in Table~\ref{Table1} independently of the chosen water model. The potential set 2 differs from the potentials used in refs~\cite{vanKlaveren2001,vanKlaveren2001b,vanKlaveren2002} only by the choice of the water model (SPC/E water was used in those references). The potential set 1, when combined with TIP4P-Ew water, coincides with the interaction potentials used in Ref~\cite{Patt2018}.

\subsection{Simulation details}

The considered clathrate structures I and II are formed of 2×2×2 unit cells taken from Takeuchi et al.\cite{Takeuchi2013}; they are made up of 368 and 1088 water molecules, respectively. The total number of cages is thus 64 (for sI) and 192 (for sII).
The GEMC simulations involved two cubic boxes. One box was initialized with an empty hydrate lattice, in a box with initial edge length 24.06 Å (for sI) or 34.62 Å (for sII); the other box was initialized with nitrogen molecules: 200 or 400 of them for simulations with the sI or sII, respectively. Periodic boundary conditions were applied separately to both boxes and the Ewald method was used to compute Coulomb interactions. The LJ and Buckingham interactions were calculated using a cutoff at 10~Å. 

Within each Monte Carlo cycle, $N$ trial moves are attempted, where $N$ equals the total number of molecules (water and nitrogen) present in both boxes. The probabilities for the various MC moves were set as follows: 1\% probability to attempt to change the volume of either the hydrate box or the \ch{N2} gas box (each having a 50\% probability of being selected); 19\% probability of attempting to transfer a nitrogen molecule from one box (chosen at random) to the other (no attempt was made to transfer a water molecule from one box to the other to avoid disrupting the clathrate structure); 40\% probability of attempting to translate a randomly selected molecule; and, finally, 40\% probability of attempting to rotate a randomly selected molecule. For the last 2 types of moves, water molecules were chosen more often than nitrogen molecules: probability 73.1\% to select the species \ch{H2O} versus 26.9\% for \ch{N2}. The maximum displacements (volume, translation, rotation) were updated periodically, each 50 MC cycles, to achieve an acceptance ratio of 50\% for each type of move.

The open-source MCCS Towhee simulation program was modified to allow to calculate conveniently the individual large and small cage occupancies. This was done by writing into a specific file, whenever a nitrogen molecule would enter or leave the hydrate box, the coordinates of the molecule, the type of move (in or out), the number of MC steps since the start of the simulation and the current hydrate box length. An ad-hoc analysis script was then used to deduce from this file and from the initial positions of all molecules, the number of \ch{N2} molecules in each type of cages as function of the MC ``time''. The simulations were run for 50,000 Monte Carlo cycles, giving a total of \num{28400000} and \num{74400000} MC steps for the simulations with sI and sII, respectively. The number of guest molecules inside the clathrate was found to have reached equilibrium after around \num{1300} Monte Carlo cycles. The uncertainties on the calculated cage occupancies are below 1\%. Some additional simulations were run by starting from a fully occupied clathrate, with all large cages singly occupied ($\theta_{\rm L} = 100\%$) or doubly occupied ($\theta_{\rm L} = 200\%$). The results were found to be identical after the equilibration period.

\section{Results and discussion}

The cage occupancies calculated from our GEMC simulations are compared with experimental data first for conditions where no double occupancy of cages has been observed experimentally (sections 3.1 and 3.2), and then for conditions where some large cages were found to be doubly occupied (section 3.3). The results presented in the next two sections pertain to the simulations of a flexible clathrate with sII with water-water interactions represented by the TIP4P-Ew water model. The results in the case of the TIP4P/Ice model will be discussed in section~3.3.

\subsection{Occupancies at 200 bar}
\begin{figure}
\includegraphics[scale=0.57]{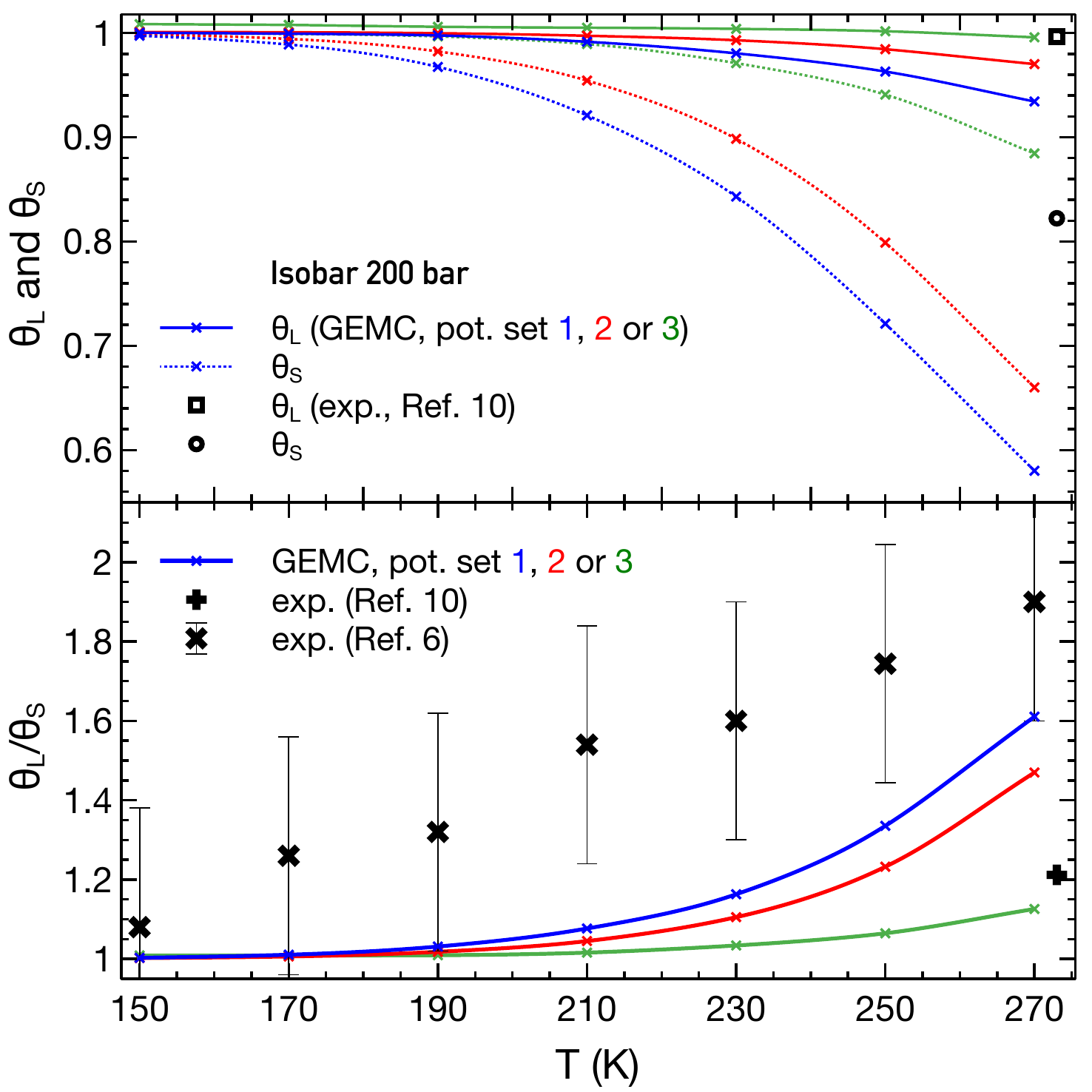}
\caption{\small Cage occupancies in large and small cages (upper panel), and their ratio (lower panel), along isobar \SI{200}{bar} according to our GEMC simulations with the TIP4P-Ew water model for the 3 sets of potentials defined in Sct.~\ref{Sct2.1}  (color code: blue, red, green for potential sets 1, 2 and 3 respectively). Experimental points from refs.\cite{Chazallon2002,Petuya2018} are also shown.
}
\label{Fig3}
\end{figure}

The results of our GEMC simulations along the isobar \SI{200}{bar} for a (flexible) clathrate with sII are shown in Fig.~\ref{Fig3}, together with experimental points for the cage occupancy ratio $\theta_{\rm L}/\theta_{\rm S}$ determined by Raman diffraction\cite{Petuya2018} and with the separate occupancies $\theta_{\rm L}=99.6\pm0.6$\% and $\theta_{\rm S}=82.2\pm0.3$\% at \SI{273}{K} determined by neutron diffraction \cite{Chazallon2002}. 
The simulations predict a preferential filling of the large cages versus the small ones in the considered temperature range, in agreement with the experimental data.

The small cages become significantly less occupied when the temperature approaches the dissociation temperature. In that region, the predicted small cage occupancies depend quite strongly on the choice of the interaction potentials, making it difficult to obtain quantitatively accurate predictions, especially for $\theta_{\rm S}$. On another hand, this sensitivity makes the \ch{N2} hydrate a particularly interesting system for testing and refining the interactions potentials employed in the simulations. The \ch{N2-N2} interactions between molecules in {\it different} cages are almost identical when using the Etters and TraPPE force fields. As the cages are here only singly occupied, the sensitivity of the predicted cage occupancies arises from the \ch{N-O} interactions. Cages are found to be more filled on average when switching from the potential set ${\color{blue}1}\rightarrow {\color{red}2}$, despite the latter potential being more repulsive at short distances ($r<3.5$\,\AA): the  ``affinity'' for filling the cages is increased by the more attractive \ch{N-O} interactions at larger distances. The increase in affinity is naturally more pronounced when switching from potential set ${\color{blue}1}\rightarrow {\color{darkgreen}3}$, since the latter potential is even more attractive when $r>3.5$\,\AA.

The cage fillings depend, in a non-linear way, on the \ch{N-O} repulsion at short distance, which determines the available cage volume, and on the binding energy $\epsilon_{\ch{N-O}}$ at the minimum of the potential well, which determines the energetic attractiveness of a cage.
The sensitivity to the binding energy $\epsilon_{\ch{N-O}}$ can be understood by using the celebrated model of van der Waals-Platteuw\cite{vdWP}, which assumes rigid cages. In this model, the cage occupancies $\theta_{\rm S}$ and $\theta_{\rm L}$ are given by a Langmuir-type formula
\begin{equation}	\label{Langmuir}
\theta_t = \frac{C_t(T) f}{1+C_t(T) f}
\end{equation}
with Langmuir constant for a cage of type $t$ ($t$=small or large)
\begin{equation}	\label{C(T)}
C_t(T) = \frac{1}{k_{\rm B} T} \int_{V_{\rm cage}}d\vec{r}\int d\vec{\Omega}\, e^{- E^{\rm guest}_t(\vec{r},\vec{\Omega})/(k_{\rm B} T)}.
\end{equation}
Here $f$ is the fugacity of the gas and $E^{\rm guest}_t(\vec{r},\vec{\Omega})$ is the potential energy of a guest molecule with orientation $\vec{\Omega}$ located at $\vec{r}$ inside a cage of type $t$.
A small shift in the binding energy $\epsilon_{\ch{N-O}}$ is demultiplied in the potential energy $E^{\rm guest}_t(\vec{r},\vec{\Omega})$ because it affects all interactions of the two \ch{N} atoms with the nearby water molecules (small and large cages in a sII hydrate are made up of 20 and 28 water molecules, respectively). If the \ch{N-O} interaction energies in a filled cage are all increased on average by \SI{5}{K}, the Langmuir constant for  a small cage will increase at \SI{270}{K} by the factor $\exp(5\cdot20\cdot2/270)\approx 2$.

The preferential filling of the large cages over the small ones can be understood on the basis of eq.~\eqref{C(T)} for the Langmuir constant. For a small molecules like \ch{N2} (diameter $4.2$\AA), the potential energy $E_{{\rm cage\ S}}(\vec{r},\vec{\Omega})$ inside a small cage is lower, i.e. more binding, on average than in a large cage because of the smaller size of the cage (the small and large cages in sII have average diameters 7.82 and 9.46$\,$\AA, respectively). In the limit of low temperatures, the Langmuir constant for small cages will therefore be larger than the one for large cages ($C_{\rm S} > C_{\rm L}$), leading to the small cages being more occupied than the large cages. As the temperature is increased, the factor $\exp(-(E^{\rm guest}_{\rm S} - E^{\rm guest}_{\rm L})/(k_{\rm B} T))$ in the ratio $C_{\rm S}/C_{\rm L}$ of Langmuir constants becomes less favorable to the small cages and the integration over the cage volume in eq.~\eqref{C(T)} favors the large cages. In other words, a small molecule like \ch{N2} will prefer the small cages at low temperature because it minimizes the energy $U$ of the system (the entropy plays no role at sufficiently low temperature), whereas it will prefer the large cages at higher temperatures because it increases the entropy $S$ of the system, the equilibrium macroscopic state being the one that minimizes the free energy $U-TS$. A preferential filling of small cages has indeed been observed in simulations at low temperatures~\cite{Patt2018}. 

It is interesting to note that the large cages are almost fully occupied along the isobar 200~bar according to our simulations. The Raman spectroscopy measurements of the ratio $\theta_{\rm L}/\theta_{\rm S}$ inform us therefore here directly on the occupancy of the small cages. A partial double occupancy of the large cages at low temperatures along this isobar cannot however be excluded, as discussed in Sct.~\ref{S2x}.  

At \SI{273}{K}, there is a quite large difference between the neutron diffraction data of Ref.~\cite{Chazallon2002} (whose uncertainty is of the order of the symbol size) and the Raman spectroscopy data of Ref.~\cite{Petuya2018}. The data of Ref.~\cite{Chazallon2002} pertains to a deuterated clathrate, but it is known that deuteration has a negligible impact on the cage occupancies, as established in Ref.~\cite{Hansen2016}.  Due to the sensitivity of the calculated occupancies on the choice of the interaction potentials, this experimental discrepancy cannot be fully lifted from our GEMC simulations at 200~bar. However, in view of the occupancies predicted when using the TIP4P/Ice model and of comparisons along isotherm 273~K (see Sct.~\ref{S2x}), and also because we expect the potential set 3 to be more accurate, the simulations appear to agree better with the lower value of  $\theta_{\rm L}/\theta_{\rm S}$ measured in Ref.~\cite{Chazallon2002}.

\subsection{Occupancies at 150 K}

\begin{figure}
\includegraphics[scale=0.57]{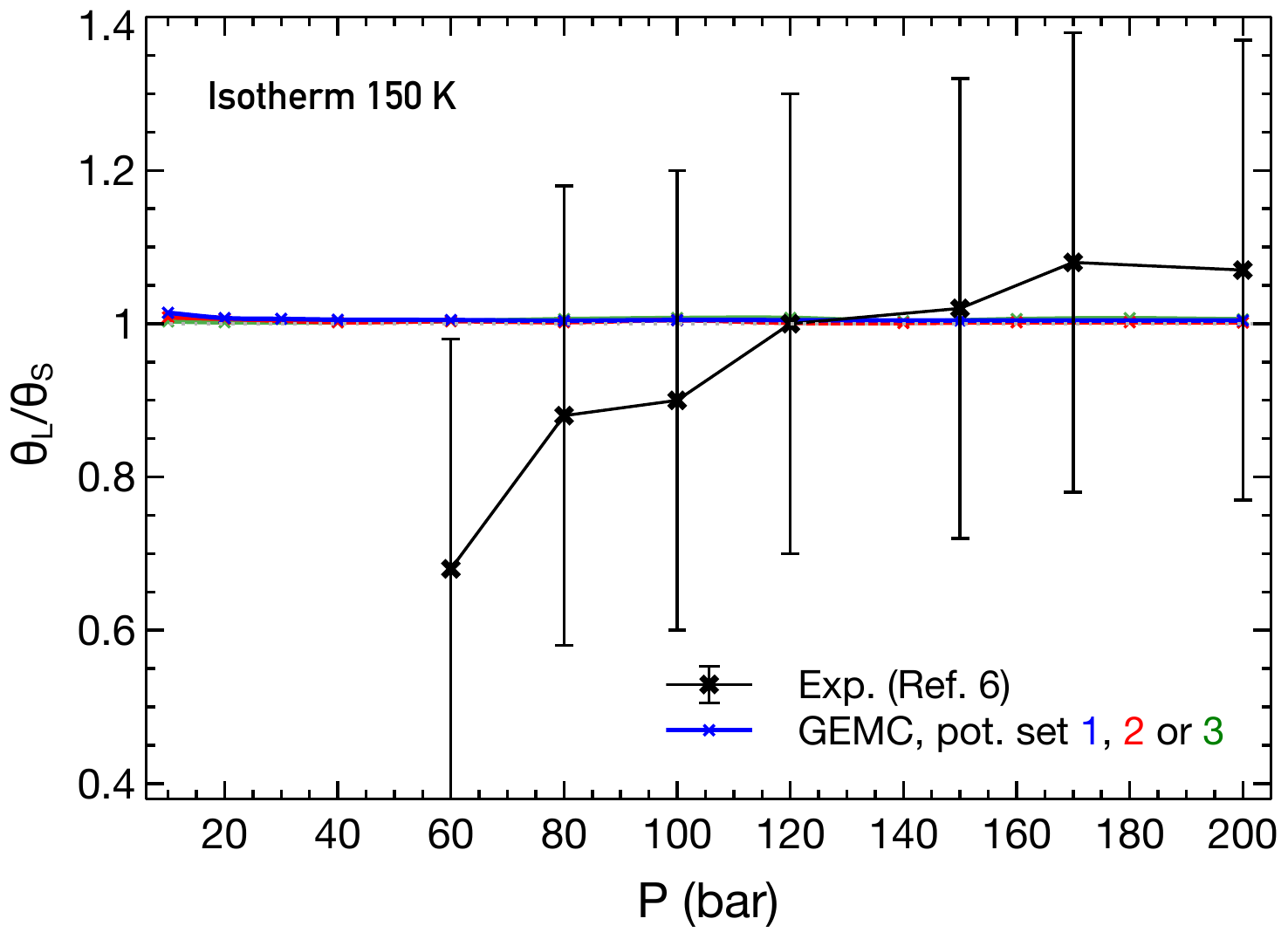}
\caption{\small 
Same as for Fig.~\ref{Fig3}, but for isotherm \SI{150}{K}.}
\label{Fig4}
\end{figure}

Occupancy ratios $\theta_{\rm L}/\theta_{\rm S}$ along the isotherm \SI{150}{K} have been determined by Raman spectroscopy. Fig.~\ref{Fig4} shows a comparison between our GEMC simulations and the corresponding experimental data\cite{Petuya2018}. The simulations indicate that all cages are fully (and singly) occupied. This can be understood as being due to the quite low temperature and to pressures that are significantly higher than the dissociation pressure $P_{\rm diss}(\SI{150}{K})\approx 3$~bar (this value is estimated by interpolating the data of Ref.~\cite{Yoon2006}). A slight preferential filling of the large cages can be seen in our simulations for pressures below \SI{20}{bar}. This agrees with an extrapolation of the isotherm \SI{150}{K} calculated, for pressures $P<\SI{10}{bar}$, by GCMC in Ref.~\cite{Patt2018}. It agrees also with the prediction $\theta_{\rm L}/\theta_{\rm S} \approx 1.001$ at 150~K and 60~bar obtained by using eq.~\eqref{Langmuir} with the Langmuir constants for \ch{N2} of Ref.~\cite{Yoon2006}.   The present prediction of a fully occupied clathrate is not sensitive to details of the interaction potentials. Taking into account the rather large error bars of the experimental points, they are compatible with our theoretical prediction $\theta_{\rm L}/\theta_{\rm S}\simeq 1$.

As explained in the previous section, a tendency for the small cages to be more occupied than the large cages, like in the experimental dataset for $P\leq\SI{100}{bar}$, is expected at low temperatures. From the simulations of Ref.~\cite{Patt2018}, the transition from $C_{\rm L}(T) > C_{\rm S}(T)$ to $C_{\rm L}(T) < C_{\rm S}(T)$ is expected to occur at around \SI{80}{K}. Even with such an ``inversion'' of the Langmuir constants, the cages are still expected to be fully occupied at the considered pressures, which are much larger than the dissociation pressure (this can be confirmed by extrapolating to higher pressures the simulation data at \SI{50}{K}, \SI{100}{K} and \SI{150}{K} of Ref.~\cite{Patt2018}). An increase of the ratio $\theta_{\rm L}/\theta_{\rm S}$ with increasing pressures could be due to some large cages encaging two molecules, as further detailed in the next section.

\subsection{Double occupancy} \label{S2x}

\begin{figure}
\includegraphics[scale=0.57]{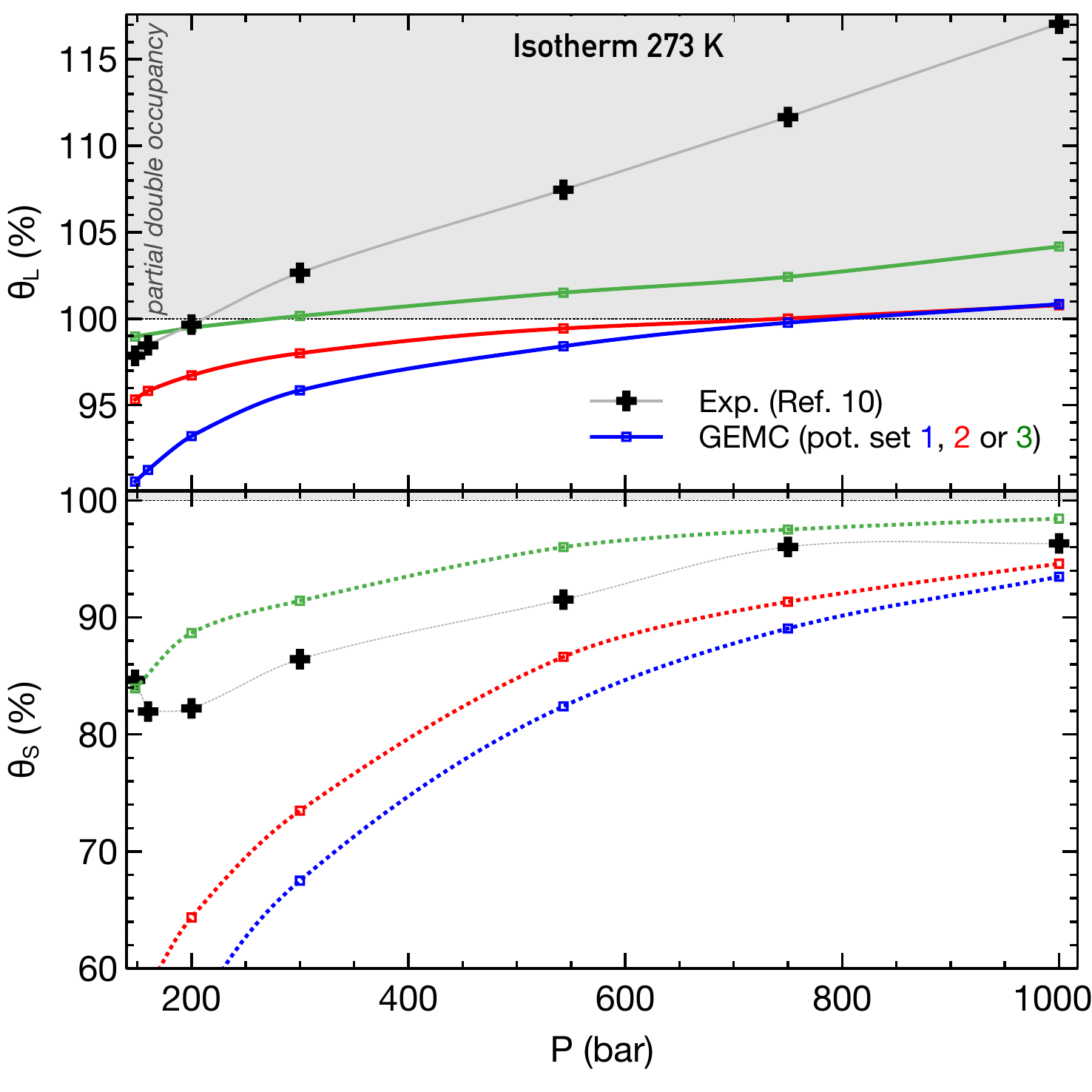}
\caption{\small Same as for Fig.~\ref{Fig3} but for isotherm 273~K (and without plot of the ratio $\theta_{\rm L}/\theta_{\rm S}$).
}
\label{Fig5}
\end{figure}

A comparison between simulation and experimental results for the cage occupancies along isotherm 273~K, from 148 bar up to the high pressure 1000 bar, is shown in Fig.~\ref{Fig5}. The potential sets 1 and 2 underestimate the large and small cage occupancies. They fail moreover to predict that some large cages in a \ch{N2} clathrate are doubly occupied at high pressures, whereas the experimental evidence for this phenomenon is strong\cite{Chazallon2002,Qin2015,Hansen2016}. The new potential set 3, introduced in Sct.~\ref{Sct2.1} by averaging an ab-initio potential energy surface, does lead to a better qualitative agreement with the experimental data: a sizable proportion of doubly occupied large cages can now be observed in those simulations at pressures above $\approx \SI{500}{bar}$.

The occupancy of a cage is sensitive to four energetic contributions: (i) the interaction energy of the guest with the surrounding water molecules, which determines the potential energy of a guest inside a cage and the available cage volume (ii) the change in water-water interaction energy upon insertion of a guest molecule (this change is due to deformations of the water network), (iii) the interaction energy between guest molecules in different cages, and (iv) the interaction energy between a guest molecule and the other guest(s) within the same cage if it is multiply occupied.
Predicting correctly multiple occupancy is more difficult than single occupancy because it is determined by a more delicate energetic balance. When inserting a second \ch{N2} molecule in a cage, the attractive energy contribution (i) remains similar than for single occupancy, but the repulsive energy (ii) becomes larger and the new repulsive energy (iv) enters into the balance.

The model of a rigid clathrate, which is employed in the van der Waals-Platteuw model\cite{vdWP}, neglects the energy cost (ii) and the fluctuations of the cage framework due to the finite temperature.  Fig.~\ref{Fig6} shows the calculated cage occupancies in simulations of a rigid \ch{N2} clathrate. In these simulations, only the nitrogen molecules can move, the water molecules are kept fixed in a perfect crystalline arrangement (no volume change of the clathrate box is furthermore attempted). 
The cages are found to be more occupied in the rigid clathrate approximation, similarly to the case of methane hydrates\cite{Wierzchowski2007,Henley2015}. Remarkably, the simulations with a rigid clathrate predict, when using our potential set 3, a sizable amount of doubly occupied large cages not only at high pressures along the isotherm \SI{273}{K}, but also at moderate pressures along the isotherm \SI{150}{K} and at low temperatures along the isobar \SI{200}{bar}. The simulations for the rigid model indicate thus that the large cages in a \ch{N2} hydrate could be doubly occupied under the latter conditions. Interestingly, this occurence of double occupancy could explain the increase of the ratio $\theta_{\rm L}/\theta_{\rm S}$ with pressure seen at \SI{150}{K} in Ref~\cite{Petuya2018} (see Fig.~\ref{Fig4}).

\begin{figure}
\includegraphics[scale=0.57]{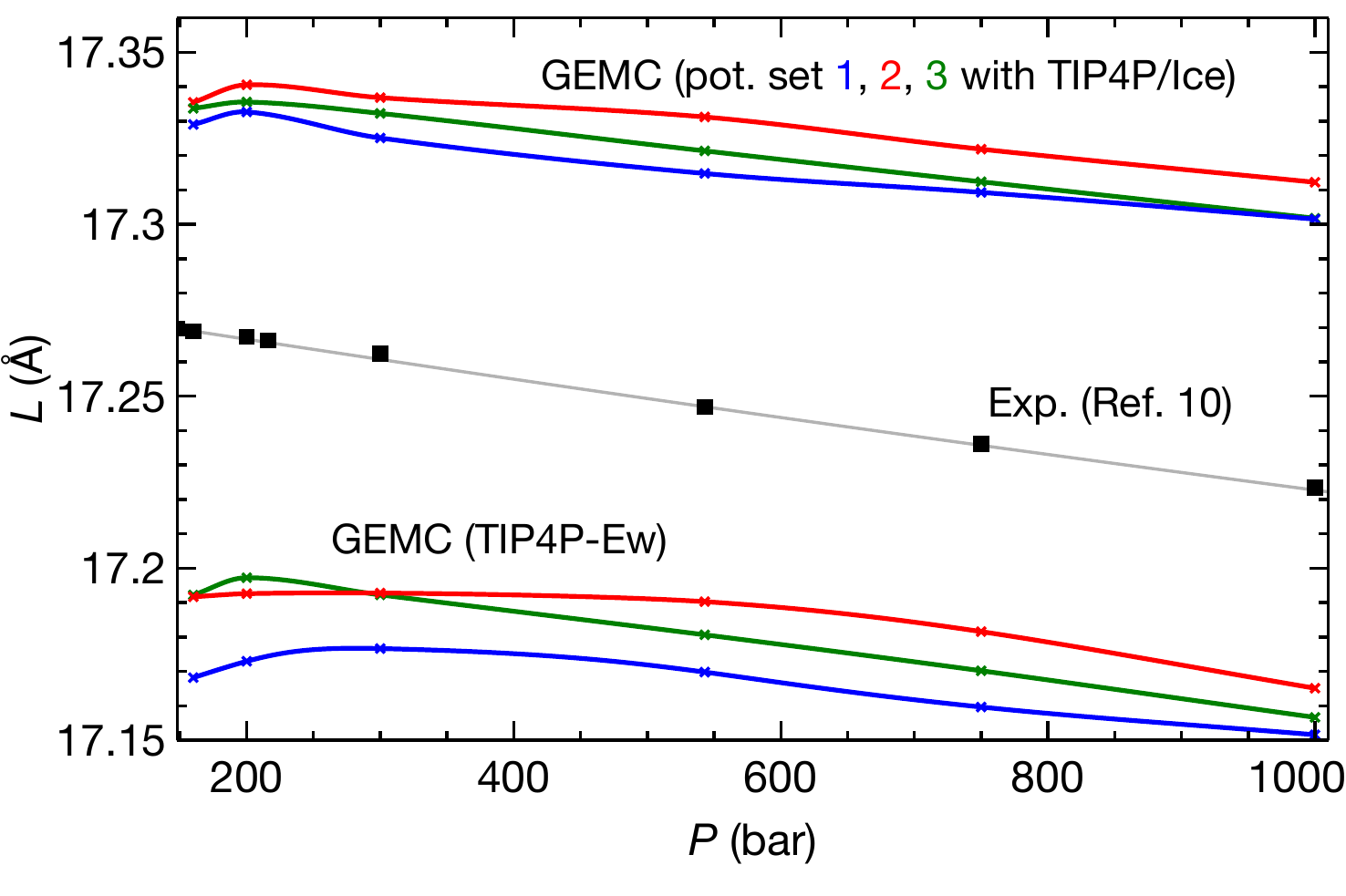}
\caption{\small Cubic lattice constant of a \ch{N2} clathrate with sII according to our simulations (colored lines, see key in plot) and to the experimental data of Ref.~\cite{Chazallon2002} (black squares).
}
\label{FigL}
\end{figure}

\begin{figure*}
\includegraphics[scale=0.6]{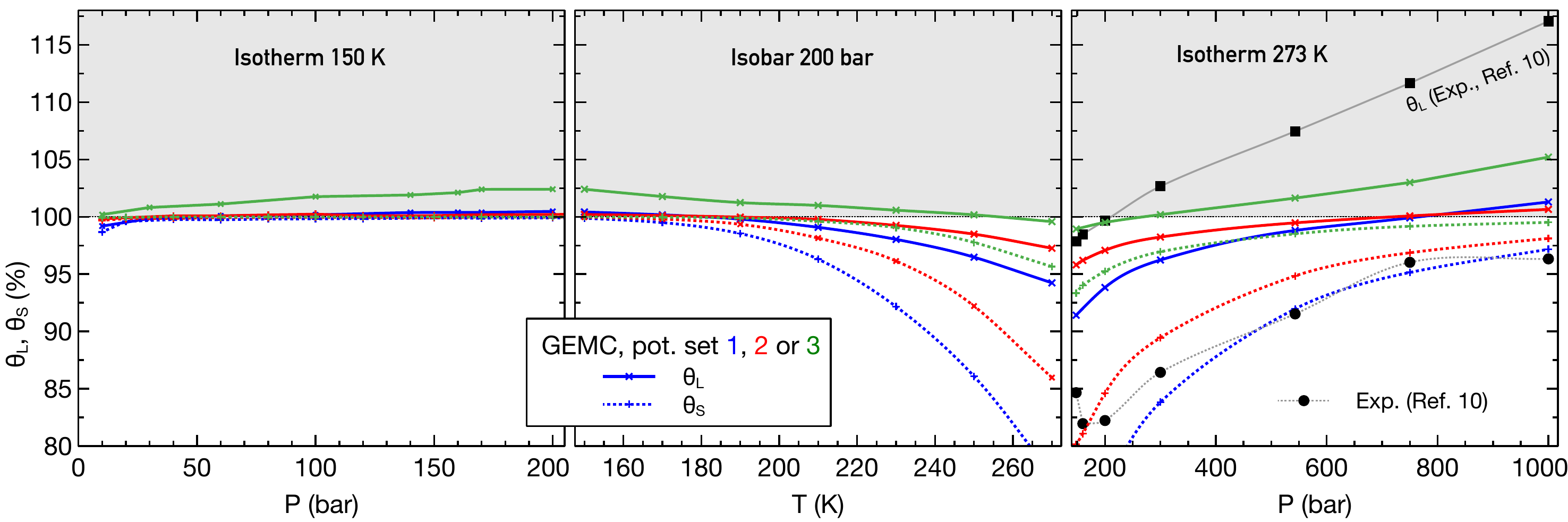}
\caption{\small Cage occupancies in large (solid lines) and small (dotted lines) cages, along isotherm 150~K (left), isobar \SI{200}{bar} (center) and isotherm 273~K (right) in a {\bf rigid} sII \ch{N2} hydrate according to our GEMC simulations with the TIP4P-Ew water model for the 3 sets of potentials. Experimental points at 273~K from Ref.\cite{Chazallon2002} are also shown.}
\label{Fig6}
\end{figure*}

\begin{figure*}
\includegraphics[scale=0.6]{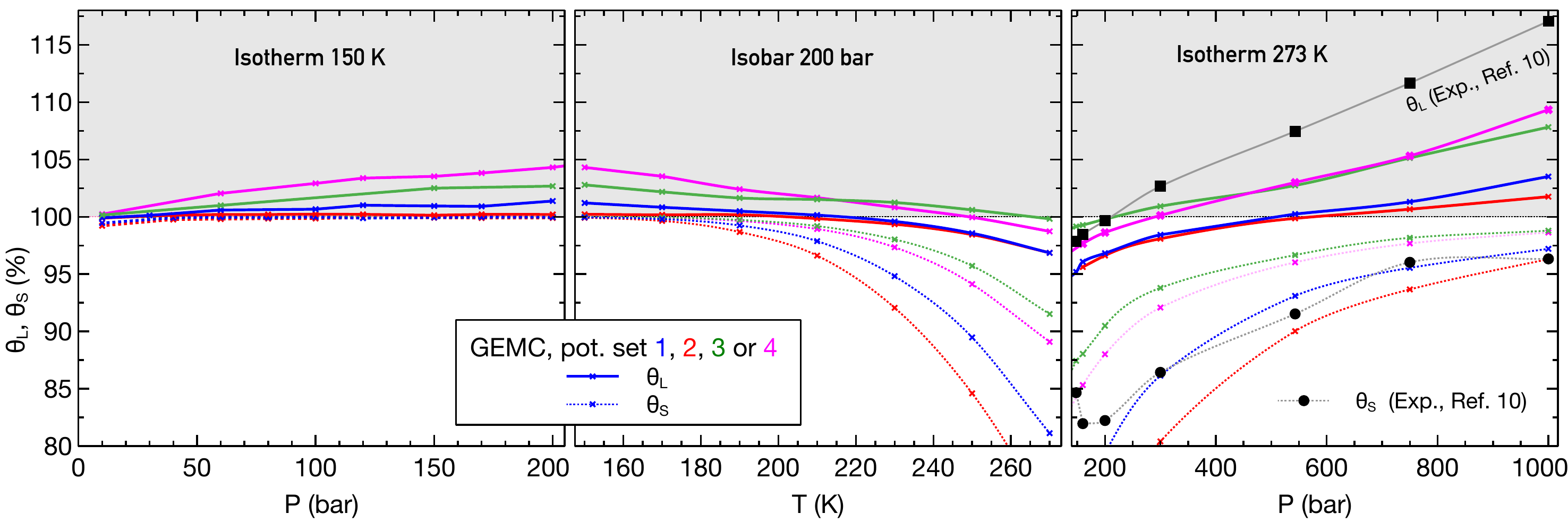}
\caption{\small Same as for Fig.~\ref{Fig6} but for GEMC simulations of a flexible \ch{N2} clathrate with the TIP4P/Ice water model. See text for the definition of the potential 4.}
\label{Fig7}
\end{figure*}

To check the sensitivity of the results on the model for water-water interactions, we have performed also simulations with the TIP4P/Ice water model\cite{TIP4Pice}. As shown in Fig.~\ref{Fig7}, the occupancies calculated in a flexible clathrate when using with TIP4P/Ice model are higher than when using the TIP4P-Ew model. This could be due to the cage framework being more rigid when modelled with TIP4P/Ice  because this model features stronger water-water interactions (12\% higher partial charges on the interaction sites and 30\% higher binding energy for the \ch{O} site – \ch{O} site interaction). Fig.~\ref{FigL} shows the lattice constant $L$ along isotherm 273~K in our simulations, for the two considered water models. The lattice constant is somewhat too small when using TIP4P-Ew, whereas it is somewhat too large when using TIP4P/Ice. The lattice spacing $L$ decreases with increasing pressure almost linearly, as in the experimental data, except at low pressures where the simulations show a small increase of $L$ with pressure when $\theta_{\rm S}$ is small. This can be attributed to an expansion of the clathrate caused by an increased filling of the cages, which more than compensates the mechanical compression.

By opposition to our previous calculations with the TIP4P-Ew model, the cage occupancies are lowered when switching from potential sets $1\rightarrow 2$ when using the TIP4P/Ice model. This is due to the use of the Lorentz-Berthelot mixing rules in the potential set~1: the binding energy $\epsilon_{\ch{N-O}}$ is higher when associating the TraPPE force field with the TIP4P/Ice model than with the TIP4P-Ew model (see Table~\ref{Table1}). The fact that both $\theta_{\rm L}$ and $\theta_{\rm S}$ are underestimated along the isotherm 273~K, whereas the potentiel set 1 provides a better agreement, especially for $\theta_{\rm S}$, confirms our finding from Sct.~\ref{Sct2.1} that the binding energy $\epsilon_{\ch{N-O}} = \SI{54.3}{K}$ associated with the combination of the TraPPE force field with the TIP4P-Ew model is too low.

The potential sets 1 and 2 associated with TIP4P/Ice water predict a non-negligible proportion of doubly occupied large cages at 273~K and 1000~bar, but the magnitude of this multiple occupancy is largely underestimated.
With the potential set 3, the occupancy of the large cages is in much better agreement with the experimental data, and predicts correctly that multiple occupancy starts at $\approx$300~bar. The occupancy of the small cages is somewhat overestimated (by about 10\% at 300 bar). Interestingly, the simulations with the TIP4P/Ice model predict, as expected from the rigid clathrate simulations (Fig.~\ref{Fig6}), that some large cages are doubly occupied at \SI{150}{K} and \SI{200}{bar}. 
As the double-occupancy Raman signature could not be disentangled in the experiments of Ref.~\cite{Petuya2018}, a verification of this prediction via a neutron diffraction experimental would be useful to further confirm the ability of GEMC simulations with the potential set 3 to correctly describe the doubly occupancy effect.

Eventually, we note that a further refinement of the parameters for the \ch{N-O} interactions can be performed to improve the agreement with the experiments at 273~K. A Buckingham potential with parameters $\epsilon=\SI{66.2}{K}$, $\gamma = 13.2$ and $r_0 = 3.24$ (i.e. an interaction that is more attractive at short distances and more repulsive at large distance than the potential 3) provides a slightly better agreement along isotherm 273~K (see magenta curve labelled ``Pot. set 4'' in Fig.~\ref{Fig7}). It predicts also a higher proportional of doubly occupied cages at 200~bar and 150~K.

%
%

\section{Conclusions}

We have compared the predictions of Gibbs ensemble Monte Carlo simulations with experimental data for the cage occupancies in \ch{N2} hydrates to assess the accuracy of such simulations, to refine the effective \ch{N-O} interaction potential beyond the common Lorentz-Berthlot mixing rules, and to help interpret cage occupancy ratios $\theta_{\rm L}/\theta_{\rm S}$ that have been recently measured  by Raman spectroscopy. Different sets of interaction potentials for \ch{N2-N2}, \ch{N2-H2O} and \ch{H2O-H2O} interactions have been considered.

The predicted cage fillings are sensitive to the choice of the interaction potentials when the filling differs significantly from 100\%. This renders accurate predictions difficult, but allows for instructive comparisons with experiments to refine the effective potentials employed. Among the considered sets of interaction potentials, some of them are not able to reproduce the known experimental fact that some large cages are doubly occupied at 273~K and high pressures.

We have found, both from comparisons with experimental data and from an analysis for an ab-initio \ch{N2-H2O} potential energy surface, that the binding energy at the bottom of the potential well for the \ch{N-O} interaction is not attractive enough when combining the TraPPE force field for \ch{N2} with the TIP4P-Ew model for water. 
The \ch{N-O} interaction potential~2 (see Table~\ref{Table1}) is more binding, but it is too repulsive at short distances for the large cages to become doubly occupied. 
By averaging an ab-initio potential energy surface of \ch{N2-H2O}, we have introduced a new effective Buckingham potential for \ch{N-O} interactions. That interaction leads to more accurate predictions for the cage occupancies, especially for the large cages. The simulations with the TIP4P/Ice water model appear to agree somewhat better with experiments than those with the TIP4P-Ew water as far as cage occupancies are concerned. Additional theoretical works on the interactions involved in \ch{N2} hydrates would naturally be useful to further improve the accuracy of GEMC simulations for this system. 

Some experimental points for the same conditions but from different experiments do not agree well with each other (for instance the points of Ref.~\cite{Chazallon2002} and Ref.~\cite{Petuya2018} at 200~bar and  $T\approx270$~K, see Fig.~\ref{Fig3}). 
The simulations with the TIP4P/Ice water model (and those within the rigid clathrate model) show that some large cages can be doubly occupied at 150~K and 200~bar. This double occupancy effect could explain the increase of the ratio $\theta_{\rm L}/\theta_{\rm S}$ measured in Ref.~\cite{Petuya2018} (see Figs.~\ref{Fig4} and \ref{Fig7}). An experimental confirmation with a separate measurement of $\theta_{\rm L}$ and $\theta_{\rm S}$ under these conditions would be useful. Experimental values for cage occupancies are most instructive in regimes where the small and/or the large cage occupancies deviate significantly from 100\%, i.e. close to the dissociation pressure (low occupancy) or under conditions where some large cages are doubly occupied (high pressures or intermediate pressure and low temperatures). Such measurements provide stringent tests for the simulations and enable refinements of the employed effective potentials.

\begin{acknowledgement}
This paper falls in the frame of the MI2C project funded by the French “Agence Nationale de la Recherche” (project ANR-15CE29-0016). The calculations were run on computers of the Institute UTINAM of the University de Franche-Comté, supported by the Région Franche-Comté and the Institut des Sciences de l’Univers (INSU), and also on the supercomputer facilities of the Mésocentre de calcul de Franche­-Comté.
\end{acknowledgement}


\begin{small}
  \providecommand{\latin}[1]{#1}
  \makeatletter
  \providecommand{\doi}
    {\begingroup\let\do\@makeother\dospecials
    \catcode`\{=1 \catcode`\}=2 \doi@aux}
  \providecommand{\doi@aux}[1]{\endgroup\texttt{#1}}
  \makeatother
  \providecommand*\mcitethebibliography{\thebibliography}
  \csname @ifundefined\endcsname{endmcitethebibliography}
    {\let\endmcitethebibliography\endthebibliography}{}

\end{small}

\newpage
\mbox{}
\newpage

\section*{Graphical TOC entry}

\begin{figure}
\includegraphics[scale=0.7]{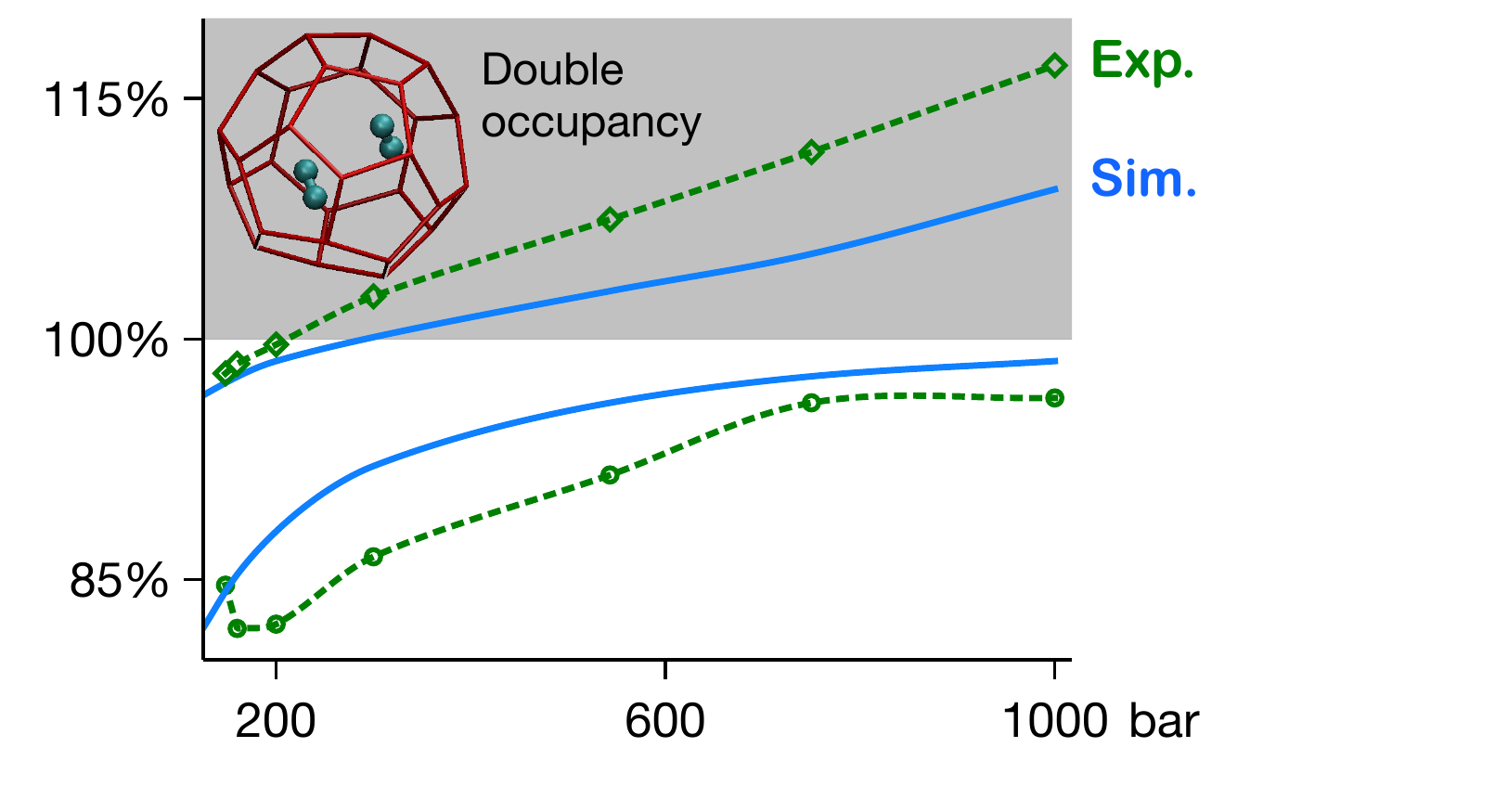}
\caption*{\small Doubly occupied large cages in a \ch{N2} clathrate hydrate}
\end{figure}

\end{document}